\documentclass[aps,pra,twocolumn,10pt,showpacs]{revtex4-1}

\usepackage{graphicx}
\usepackage{bm}
\usepackage{amsthm,amsfonts,amstext,amsmath,amssymb}
\usepackage{latexsym}
\usepackage{hyperref}
\usepackage[margin,index]{fixme}
\usepackage[utf8]{inputenc}
\usepackage[makeroom]{cancel}
\usepackage[T1]{fontenc}
\usepackage[usenames,dvipsnames]{color}

\newcommand{\be}{\begin{equation}}
\newcommand{\ee}{\end{equation}}

\newcommand{\id}{\boldsymbol 1}

\newcommand{\ket}[1]{|#1\rangle}
\newcommand{\av}[1]{\left\langle #1\right\rangle}

\begin{document}
\title{Bose-Hubbard model with random impurities: 
Multiband and nonlinear hopping effects}

\author{Julia Stasi\'nska$^\heartsuit$, Mateusz \L{}\k{a}cki$^\diamondsuit$, Omjyoti Dutta$^\diamondsuit$, Jakub Zakrzewski$^{\diamondsuit,\clubsuit}$, and Maciej Lewenstein$^{\heartsuit,\spadesuit}$}
\affiliation{
\mbox{$^\heartsuit$ ICFO-Institut de Ciencies Fotoniques, Mediterranean Technology Park, 08860 Castelldefels (Barcelona), Spain}
\mbox{$^\diamondsuit$ Instytut Fizyki imienia Mariana Smoluchowskiego, Uniwersytet Jagiello\'nski, \L{}ojasiewicza 11, 30-348 Krak\'ow, Poland}
\mbox{$^\clubsuit$ Mark Kac Complex Systems Research Center, Jagiellonian University,   \L{}ojasiewicza 11, 30-348 Kraków, Poland}
\mbox{$^\spadesuit$ ICREA-Instituci\`o Catalana de Recerca i Estudis Avan\c{c}ats, E08011 Barcelona, Spain}}

\date{\today}

\begin{abstract}
We investigate the phase diagrams of theoretical models describing bosonic atoms in a lattice in the presence of randomly localized impurities. By including multiband and nonlinear hopping effects we enrich the standard model containing only the chemical-potential disorder with the site-dependent hopping term. We compare the extension of the MI and the BG phase in both models using a combination of the local mean-field method and a Hartree-Fock-like procedure, as well as, the Gutzwiller-ansatz approach. We show analytical argument for the presence of triple points in the phase diagram of the model with chemical-potential disorder. These triple points however, cease to exists after the addition of the hopping disorder.
\end{abstract}
\maketitle


\section{Introduction}

Ultracold atoms in optical lattices serve now as a routine tool to study various lattice models derived from other areas of physics, such as condensed matter or high energy physics. They often enrich the original models with additional features accessible 
due to extreme controllability and versatility of possible experimental realizations (for recent reviews see
\cite{Lewenstein07,Bloch08,Lewenstein12}). This potential of cold atoms was recognized for the first time thanks to a seminal theoretical proposal of Jaksch {\it et al.} \cite{Jaksch98}. Soon followed the experimental demonstration of the superfluid-Mott insulator phase transition \cite{Greiner02}, and the intensive studies of the effects of disorder on cold-atom systems \cite{Damski03,Roth03}. After a series of attempts \cite{Lye05,Fort05,Schulte05,Clement05,Clement06,Schulte06,Lye07} the Anderson localization for non-interacting (expanding) atoms was unambiguously demonstrated in \cite{Billy08}, 
The experimental studies of the phase diagram of interacting bosonic atoms in a disordered potential revealed the long-debated gapless insulating Bose glass (BG) phase \cite{Fallani07}, stimulating at the same time the discussion on the ways to observe and detect the BG phase \cite{Roux08,Roscilde08,Zakrzewski09,Delande09,Roux13}. Various aspects of Anderson localization in cold atoms are reviewed in Ref. \cite{Aspect09,Modugno10}, while the importance of disorder studies in a broader context of quantum simulators is presented in \cite{LSPLew10}.

Optical implementations of quenched disorder are unique in the sense that the disorder can be, in principle, controlled with high precision on demand. Various methods of creating disorder have been proposed. A speckle pattern collimated on the atomic sample \cite{Damski03} gives rise to a truly random intensity landscape, which follows the exponential distribution.
Another proposed scheme \cite{Damski03,Roth03} applies several
(at least two) laser fields with different frequencies. For an appropriate choice of frequencies the resulting potential is quasi-periodic, and for a finite sample hardly distinguishable from a truly random case (see \cite{Roati08}, and also \cite{Orso09,DErrico14,Tanzi13} for recent results).

Yet another interesting way to create the disordered potential is to use the interactions between atoms. By pinning a secondary type of atoms in an optical potential we obtain the disorder with binary (or Bernoulli) distribution.
Such proposal, originating from the work of Gavish and Castin \cite{Gavish05}, has been frequently discussed theoretically \cite{Massignan06,Mering2008pra,Krutitsky2008pra,Buonsante2009pra,
Stasinska2012njp}, and only recently implemented experimentally \cite{Gadway2011prl} (note that in the early papers \cite{Ospelkaus2006prl} the impurities were mobile). This is the type of disorder we consider in this paper.

%

Specifically, we study the bosonic atoms in the optical lattice potential, which interact with immobile randomly distributed atoms of a secondary (fermionic) species. Such a situation is routinely described using the Bose-Hubbard model with the random potential
\begin{equation}\label{eq:Hamiltonian_1}
H_{1}=-t\sum_{\langle i,j \rangle} b_i^{\dagger} b_j
+\frac{U}{2} \sum_i b_i^{\dagger}b_i^{\dagger} b_i b_i-\sum_i \left(\mu-\gamma\omega_i\right) b_i^{\dagger}b_i,
\end{equation}
where $b_i, b_i^{\dagger}$ are the bosonic annihilation and creation operators, $t$ is the tunneling, the interaction constant is denoted by $U$, $\mu$ is the chemical potential and $\langle i,j \rangle$ denotes the summation over the nearest neighbors. The parameter $\gamma$ characterizes the strength of the disorder and $\omega_i$ is a random variable  with binary distribution (taking value $1$ with probability $p$ when the heavy background fermion is present at the $i$th site, and $0$ with probability $(1-p)$, when no impurity is present). This Hamiltonian was already studied in several works \cite{Mering2008pra,Krutitsky2008pra,Buonsante2009pra}, where the emergence of the Mott insulating (MI) phase with non-integer filling related to the impurity density was demonstrated, and the MI phase was shown to survive for arbitrarily strong disorder unlike in the continuous disorder case.
The Bose-Hubbard model with different forms of  diagonal disorder has been reviewed in \cite{Pollet13} while random on-site interactions have been considered in \cite{Gimperlein05}

It has been shown recently that the simple Bose-Hubbard description for fermion-boson mixture may not be adequate for stronger interspecies interactions. The shift of the observed transition between the superfluid (SF) and the MI phases, observed experimentally in Refs. \cite{Ospelkaus2006prl,Gunter06,Best09} (for the analogous effects with Bose-Bose mixtures and tightly trapped bosons, see \cite{Thalhammer2008,Will2010}), could not be explained with this simple description. It has soon been realized that density-dependent tunneling terms as well as contributions coming from higher Bloch bands are necessary to describe the systems in question
\cite{Luhmann08,Mering2011pra,Jurgensen12,Bissbort2012pra}, whenever the interspecies interaction becomes strong enough. As a consequence, for bosons interacting with immobile fermionic impurities the disorder affects also the tunneling. While similar corrections could be also taken into account for boson-boson interactions, we do not include them to simplify the picture, assuming  boson-boson interactions to be  sufficiently weak  (see a recent review \cite{review} for discussion of different possible contributions). In this simplified picture we  add to the Hamiltonian (\ref{eq:Hamiltonian_1}) the term $T(\omega_i+\omega_j)] b_i^{\dagger} b_j$ yielding the tunneling dependent on the presence of the heavy fermion. The density induced tunneling coefficient $T$, proportional to boson-fermion interaction strength,  seems at first independent of the standard tunneling $t$. However, this is not the case in the optical lattice potential, where both $t$ and $T$ depend on the potential depth (when proper 
Wannier functions are used to evaluate them), and are (see e.g. \cite{review}) approximately proportional to each other for standard depths of optical lattices. 
Thus, we may assume $T=\alpha t$
obtaining the Hamiltonian:
\begin{eqnarray}\label{eq:Hamiltonian_2}
H_{2}&=&-t\sum_{\langle i,j \rangle} [1+\alpha(\omega_i+\omega_j)] b_i^{\dagger} b_j\\
&&+\frac{U}{2} \sum_i b_i^{\dagger}b_i^{\dagger} b_i b_i-\sum_i \left(\mu-\gamma\omega_i\right) b_i^{\dagger}b_i.\nonumber
\end{eqnarray}
Here, $\alpha$ and $ \gamma$ depend on the interaction between the two species and, typically, $|\gamma| > |\alpha|$ \cite{review}.
They are of the same sign, and below we only discuss the case of positive $\alpha$ and $\gamma$. This corresponds to the repulsive boson-fermion interaction. Note that in this model a single local random variable related to the presence of the impurity, enters in the potential and the tunneling terms of the Hamiltonian of the system. For the sake of comparison we shall also study a different model in which the disorder in the tunneling is given by independent random variables $\Omega_i\neq \omega_i$  (see also \cite{Grzybowski2006pss,Kruger2009prb} for models with simultaneous potential and hopping disorder):
\begin{eqnarray}\label{eq:Hamiltonian_3}
H_{3}&=&-t\sum_{<i,j>} [1+\alpha(\Omega_i+\Omega_j)] b_i^{\dagger} b_j\\
&& +\frac{U}{2} \sum_i  b_i^{\dagger}b_i^{\dagger} b_i b_i-\sum_i \left(\mu-\gamma\omega_i\right) b_i^{\dagger}b_i.\nonumber
\end{eqnarray}

The main goal of the present paper is to compare the phase diagrams of Hamiltonians $H_1, H_2$ and $H_3$. To this end we develop a method being a combination of a site-dependent decoupling mean-field method with a ``Hartree-Fock-like'' procedure \cite{Stasinska2014}. Independently we use also the Gutzwiller-ansatz approach. Note that the type of disorder that we study does not fulfill the assumptions of the "theorem of inclusions", valid for continuously distributed bounded disorder \cite{Pollet09}.  We perform the analysis mostly in two dimensions (2D) as the mean-field approaches cannot be regarded as accurate in 1D. Our main results are: i) for the Hamiltonian $H_1$ there is a direct MI-SF transition at the tips of the Mott lobes in the thermodynamic limit; ii) for the Hamiltonian $H_2$ the Mott lobes are smaller, and the direct MI-SF transition at their tips disappears, although the region of BG is very narrow; iii) finally, for $H_3$ there is no direct MI-SF transition and the region of BG is wider in 
comparison to $H_2$. One of the main aspects of this paper is also the use of the mean-field theory combined with the simple Hartree-Fock approach, quite different from what has been proposed so far \cite{Fisher89,Sheshadri1995prl,Buonsante2007pra,Pisarski2011pra,Niederle2013,Bissbort2009epl,Bissbort2010pra}, and quite efficient in determining the phase boundary between the BG and the SF phase.

The paper is structured as follows. In section \ref{sec:survey} we briefly recall the mean-field approaches applied in the studies of the disordered Bose-Hubbard model. In section \ref{sec:method} we introduce a method being a combination of a local mean-field approach and a Hartree-Fock-like method. The standard Gutzwiller approach used later for comparison is presented in  \ref{sec:Gutzwiller}. In section \ref{sec:results} we apply the theory developed in section \ref{sec:method} to compare the phase diagrams of the bosonic atoms interacting with immobile impurities on a lattice described by the Hamiltonians $H_1$, $H_2$ and $H_3$ and confront these results with the Gutzwiller approach.  Finally, we conclude in section \ref{sec:summary} by summarizing the obtained results.


\section{Brief survey of mean-field approaches for the disordered Bose-Hubbard model}\label{sec:survey}
The phase diagram of the disordered Bose-Hubbard model has been studied by a number of methods including the quantum Monte Carlo \cite{Krauth1991,Scalettar1991,Batrouni1992,Kisker1997,Prokofev2004}, renormalization group \cite{Singh1992,Svistunov1996}, density-matrix  renormalization group techniques \cite{Pai1996,Rapsch1999,Lee2001}, tensor networks-based algorithms, or various mean-field approaches \cite{Fisher89,Sheshadri1995prl,Damski03,Krutitsky2006}. In this work we propose an extension of the local mean-field method, thus let us first briefly review the mean-field approaches used earlier. 

The local mean-field method was introduced in Ref. \cite{Fisher89}, and  further developed in \cite{Buonsante2007pra}, where the boundary of the Mott lobe was linked to the stability of the zero solution of the self-consistency equations, which is then studied through linearization of those equations. Moreover the authors suggest that the presence of the BG surrounding the MI phase could be inferred from the spectral properties of the random matrix, which appear in the linearized problem.
In \cite{Pisarski2011pra} the authors generalize the inhomogeneous site-dependent mean-field to clusters, which allows them to include the (short-range) spatial correlations. Sheshadri and co-workers \cite{Sheshadri1995prl}  proposed to study the Bose-Hubbard model with disordered potential using the inhomogeneous generalization of the site-decoupling mean-field method, i.e., the local mean-field theory . There the hopping term is decoupled as $b_i^{\dagger}b_j\approx b_i^{\dagger}\av{b_j}+b_j\av{b_i^{\dagger}}-\av{b_i^{\dagger}}\av{b_j}$ yielding single site Hamiltonians coupled to neighbouring sites only through SF amplitudes $\psi_j=\av{b_j}$. The authors then diagonalize the local Hamiltonians in the occupation-number basis and determine self-consistently the values of $\psi_j$ minimizing the energy. The BG-SF transition is characterized through the percolation of sites with non-zero SF parameter.

A description equivalent to the local mean-field theory may be achieved by minimizing the average energy over a variational manifold composed of products of single site wave vectors. Indeed, under such assumptions $\langle a_i a_j^\dagger\rangle = \langle a_i \rangle \langle a_j^\dagger\rangle$. This numerical ansatz is called the Gutzwiller ansatz.

Yet another approach, the stochastic mean-field method, was proposed by Bissbort and Hofstetter in \cite{Bissbort2009epl} (see also \cite{Bissbort2010pra}). There, the starting point is also the decoupling of the tunneling term, however instead of choosing a different mean-field parameter for each site, the authors consider the probability distribution $P(\psi)$, reducing the description to effectively single-site problem. The probability distribution of $\psi$ is then found self-consistently to ensure the compatibility of $P(\psi)$ with the distribution on the neighbouring sites.

Finally, in a recent work Niederle and Rieger \cite{Niederle2013} compare the results obtained with the local mean-field theory and the stochastic mean-field method with the quantum Monte-Carlo results. They conclude that the identification of different phases based on averaged quantities obtained through the mean-field approaches is misleading. Instead, the authors propose to distinguish the phases through the presence and percolation of the SF clusters finding an excellent agreement with the quantum Monte-Carlo studies.

In what follows we will take the latter approach, i.e., study the percolation of the SF clusters; we propose, however,  a different method to determine the distribution of the ``superfluid particles'' in the lattice. To that end we combine the mean field approach with Hartree-Fock-like method mixing different mean field modes. In this way, at a little numerical effort, we can go beyond the standard mean field approach.

\section{Local mean-field approach combined with a Hartree-Fock-like method}\label{sec:method}

The proposed method is a compromise between the local mean-field description which, based on the product-state description of the system, may not capture correctly the long-range correlations, and the resource-demanding numerical approaches. We also make an attempt to include the spatial correlations on top of the simple local mean-field description.


Since the Hamiltonians $H_1$ and $H_2$ can be considered as special cases of the Hamiltonian $H_3$, in what follows we will concentrate on the latter. 

\subsection{Standard mean-field approach - local Hamiltonian} 

Like in the routine mean-field approach \cite{Fisher89} we begin by decoupling the Hamiltonian (\ref{eq:Hamiltonian_3}) using the standard approximation $b_i^{\dagger}b_j\approx b_i^{\dagger}\av{b_j}+b_j\av{b_i^{\dagger}}-\av{b_i^{\dagger}}\av{b_j}$ \cite{Sheshadri1993epl} and introducing a local mean-field parameter $\psi_i=\av{b_i}$. As a result we obtain:
\begin{eqnarray}
H&=&H_i+\bar{t} \sum_{\langle i,j \rangle} [1+\alpha(\Omega_i+\Omega_j)] \psi_i^{*} \psi_j,\label{eq:Hamiltonian}\\
H_i&=&-\bar{t}\sum_{\langle j\rangle_i}[1+\alpha(\Omega_i+\Omega_j)]\psi_j(b_i+b_i^{\dagger})\nonumber\\
&&+\frac{1}{2} b_i^{\dagger} b_i^{\dagger} b_i b_i-(\bar{\mu}-\bar{\gamma}\omega_i) b_i^{\dagger}b_i, \label{eq:Hamiltonian_i}
\end{eqnarray}
where we also express all the parameters in the units of $U$, i.e., $\bar{t}=t/U,\bar{\mu}=\mu/U,\bar{\gamma}=\gamma/U$.

\subsection{Standard mean-field approach - energy minimization and self-consistency} 

The ground state, or more generally Gibbs  energy of the local mean-field Hamiltonian is a highly nonlinear function of the local mean fields   $ \psi_j$'s.
The next step in the standard approach is to find a minimum of the energy under the constraint that $\av{b_j}= \psi_j$. This is in general a complicated task, but as long as we are interested in finding the boundaries of the MI phase, the analysis can be restricted to small values of  $\psi_j$'s, where the energy is a quadratic form of $\psi_j$'s.
The solutions of  the self-consistency equations can be then obtained via a perturbative expansion up to first order in $\bar t$:
\begin{equation}\label{eq:self_consistent}
\psi_i:=\av{b_i}=\sum_{\langle j\rangle_i}\bar t \mathcal{R}_{ij} \psi_j,
\end{equation}
with the random matrix
\begin{eqnarray}
&&\mathcal{R}_{ij}=[1+\alpha(\Omega_i+\Omega_j)]\\ &&\times\left(\frac{\bar{n}_i+1}{ \bar{n}_i-\bar \mu+\bar \gamma\omega_i}-\frac{\bar{n}_i}{(\bar{n}_i-1)-\bar \mu+\bar \gamma\omega_i}\right),\nonumber
\end{eqnarray}
where $\bar n_i$ is chosen is such a way that $\bar n_i-(1-\bar\gamma \omega_i) < \bar\mu < \bar n_i+\bar\gamma \omega_i$.

The MI phase corresponds to such $\bar t,\bar \mu$ that the system admits only a trivial solution (the energy has the minimum at  $\psi_j=0$ for all $j$).  Clearly, this occurs if and only if $\det(\bar t \mathcal{R}-\id)\neq 0$, in other words, whenever
\begin{equation}\label{eq:Mottbound}
\bar t \max[\lambda(\mathcal{R})]<1,
\end{equation}
where $\lambda(\cdot)$ denotes the spectrum of a matrix. Once $\bar t$ exceeds the critical value (the condition (\ref{eq:Mottbound}) is violated) we enter a phase with at least one unstable mode that attains a non-zero value, determined by the full nonlinear dependence of the energy on $\psi_i$'s; obviously, the  linear theory predicts only the instability and, formally, a value of the amplitude of the corresponding unstable mode tending to infinity.

\subsection{Non-standard mean-field approach - populating unstable modes} 

The idea here is a simple one - we consider all modes that are unstable (i.e. these that violate the inequality (\ref{eq:Mottbound})).
Note that for finite systems the spectrum of the matrix is discrete, hence only for specific values of $\bar t$ different eigenvectors of $\mathcal R$ become the solutions of Eq. (\ref{eq:self_consistent}), i.e., a fixed point of the linear map $\bar t\mathcal R$. Hence, in general it is more physical to consider the vectors $\psi$ belonging to the unstable manifold of the map $\bar t \mathcal R$, rather than the individual solutions of the self-consistency equation (\ref{eq:self_consistent}). We denote by $\mathcal Q(\bar t)$ the set of indices for the eigenvectors of $\bar t\mathcal R$ corresponding to the eigenvalues larger then one. These are the modes that we expect to be populated. 

Consequently, we define new modes, $a_k, a_k^{\dagger}$, corresponding to right and left eigenvectors of $\mathcal{R}$, $\psi^{(k)}, \overline{\psi}^{(k)}$, and related to the original modes as
\begin{equation}\label{eq:modes}
b_i=\sum_k \psi_i^{(k)}a_k, \quad  b_i^{\dagger}=\sum_k \overline{\psi_i}^{(k)}a_k^{\dagger}.
\end{equation}
and express the initial (not decoupled) Hamiltonian (\ref{eq:Hamiltonian_3}) in terms of these operators. Our aim is to minimize the energy with respect to the population $\{n_k\}$ of the new modes. Taking the ground state in the form $\ket{g.s.}=\sum_{k\in\mathcal Q (\bar t)} 1/{\sqrt{n_k!}} a_k^{\dagger}{}^{n_k}\ket{0}$ we obtain
\begin{equation}\label{eq:energy}
\av{H_3}_{\ket{g.s.}}=\sum_{k\in \mathcal Q(\bar t)} n_k E_k + \sum_{k,l\in \mathcal Q(\bar  t)} n_k(n_l-\delta_{k,l}) O_{kl},
\end{equation}
where $\mathcal Q(\bar t)$ is the set of indices defined before, and
\begin{eqnarray}\label{eq:defs}
E_k&=&-\bar t \sum_{<i,j>} [1+\alpha(\Omega_i+\Omega_j)]\overline{\psi}_k^{(i)}\psi_k^{(j)}+\sum_i \bar \gamma\omega_i\overline{\psi}_k^{(i)}\psi_k^{(i)},\nonumber\\
O_{kl}&=&\frac{1}{2}\sum_i (2-\delta_{k,l})\overline{\psi}_k^{(i)} \overline{\psi}_l^{(i)}
\psi_k^{(i)}\psi_l^{(i)}.
\end{eqnarray}
The number of particles is adjusted for each value of $\bar t$ to match the chemical potential $\bar \mu$ in the definition of $\mathcal{R}$.

Knowing the population of the modes, we determine the distribution of the number of particles $n_i$ in the lattice as:
\begin{equation}\label{eq:avnum}
\av{n_i}=\sum_k n_k \bar{\psi}_i^{(k)}\psi_i^{(k)}.
\end{equation}
As long as the regions of non-zero (i.e. of absolute value greater then a threshold value) number of particles {\color{blue} }  are disconnected we identify the phase as the BG. The boundary of this phase and the transition to the SF phase is given by $\bar t$ for which the sites with non-zero $n_i$ begin to percolate. This condition seems analogous to the approach of Niederle and Rieger\cite{Niederle2013}, however, there is an important difference.  While in \cite{Niederle2013} the percolation of mean-field occupations of sites is directly taken as a superfluid border, in our approach we rebuild these occupations from the Hartree-Fock-like procedure discussed above.

\section{Gutzwiller-ansatz}\label{sec:Gutzwiller}

The results obtained following the approach discussed above will be in the next Section confronted with the standard Gutzwiller approach. In the latter 
the minimization of the energy functional $E[\psi]=\langle \psi | H |\psi\rangle$ over the product states of the form 
\be
|\psi\rangle=\prod_i \sum_n f_i(n) |n\rangle_i,
\label{gucio}
\ee
and subject to the normalization condition $\langle \psi | \psi\rangle=1$ is performed. One needs to minimize over the expansion coefficients $f_i(n)$, with $|n\rangle_i$ denoting the Fock states at site $i$.

The numerical minimization of such a nonlinear problem is simple for a homogeneous system, as minimization variables $f_i(n)$ without loss of generality may be considered site-independent. For low densities one limits possible occupations to, say, $n_{max}=5$ making it a five-parameter minimization (taking into account the normalization). In such case the standard conjugate-gradient minimization algorithms always converge then to the global minimum. In contrast, in the presence of disorder the number of minimized parameters increases to $n_{max}L$ where $L$ is the number of sites. More importantly, the energy landscape of the energy functional $E[\psi]$ may contain plenty of local minima in the presence of disorder. To reach the (hopefully) global ground state additional precautions have to be made (starting from different initial conditions, perturbing the found minima to check whether they are the local or global ones etc.).

We have used a 2D lattice which contained $M\times M=40\times40$ lattice sites ($L=1600$) with $n_{max}=4$.  Periodic boundary conditions have been assumed. 

In the MI phase the mean field solution yields a Fock state at each site with a vanishing variance
of the occupation number $\sigma^2_i\equiv \langle (b_i^\dagger b_i)^2 \rangle - 
\langle b_i^\dagger b_i \rangle^2$. Thus a convenient criterion for the disappearance of MI is that $\sigma^2  \equiv \max \sigma^2_i$ exceeds a given threshold value $s_m$. Of course the obtained results depend to some extent on the value of $s_m$. Practically the dependence is quite small, we found 
$s_m=0.001$ leads to an almost perfect agreement between the Gutzwiller ansatz prediction and the eigenvalue condition \eqref{eq:Mottbound}.

Having the distribution of $\sigma^2_i$ one could define the BG-SF transition as a border at which non-zero $\sigma^2_i$ percolate. However, we employ another method calculating the classical property, the superfluid fraction (SF), $\rho_s$ which should vanish in BG phase. It is obtained using the ``boost'' method \cite{Lieb02,Roth03,Damski03} transferring the system to the moving frame by making the tunneling amplitude complex. Explicitly for tunneling along the $x$-axis we change $t\rightarrow t\exp(\pm i\varphi)$ for tunnelings (with sign corresponding to the direction of tunneling and $\varphi$ being a small angle). This corresponds to the presence of a constant flux proportional to $\varphi$ and the SF is obtained as \cite{Lieb02,Roth03,Damski03} $\rho_s= (E(\varphi)-E(0))/Nt\varphi^2$ where N is the total number of atoms and $E(\varphi)$ is the ground state energy at a given value of $\varphi$. In practice the parabolic dependence of  $E(\varphi)$ on $\varphi$ is tested to extract a reliable SF. Let 
us, however, mention that the SF calculated in this manner is not entirely correct in the presence of the density dependent tunneling (for a detailed discussion of superfluid fraction definition in various situations see \cite{Rousseau2014}). We believe, however, that since in our case the density dependent tunneling terms are small, the onset of the superfluidity may be well estimated by the traditional approach.

Before discussing the results let us stress that the SF fraction introduced in this way has little in common with the ``averaged SF order parameter'' criticized in  \cite{Niederle2013}. The results obtained for the superfluid border using a proper superfluid fraction reproduce, in fact, the percolation 
border of \cite{Niederle2013} with quite a good accuracy. They are shown below together with the HF percolation threshold.

\section{Results}\label{sec:results}
%
\begin{figure}[!ht]
\includegraphics[scale=0.5,trim=21 0 0 0]{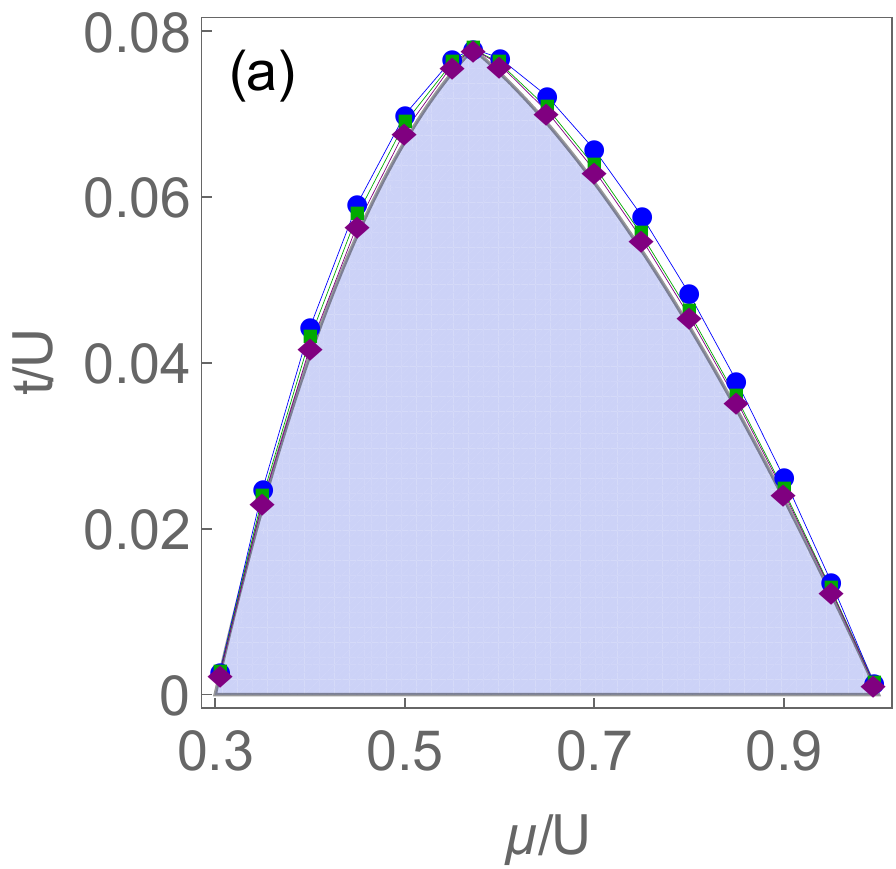}\includegraphics[scale=0.5,trim=16 0 0 0,clip=true]{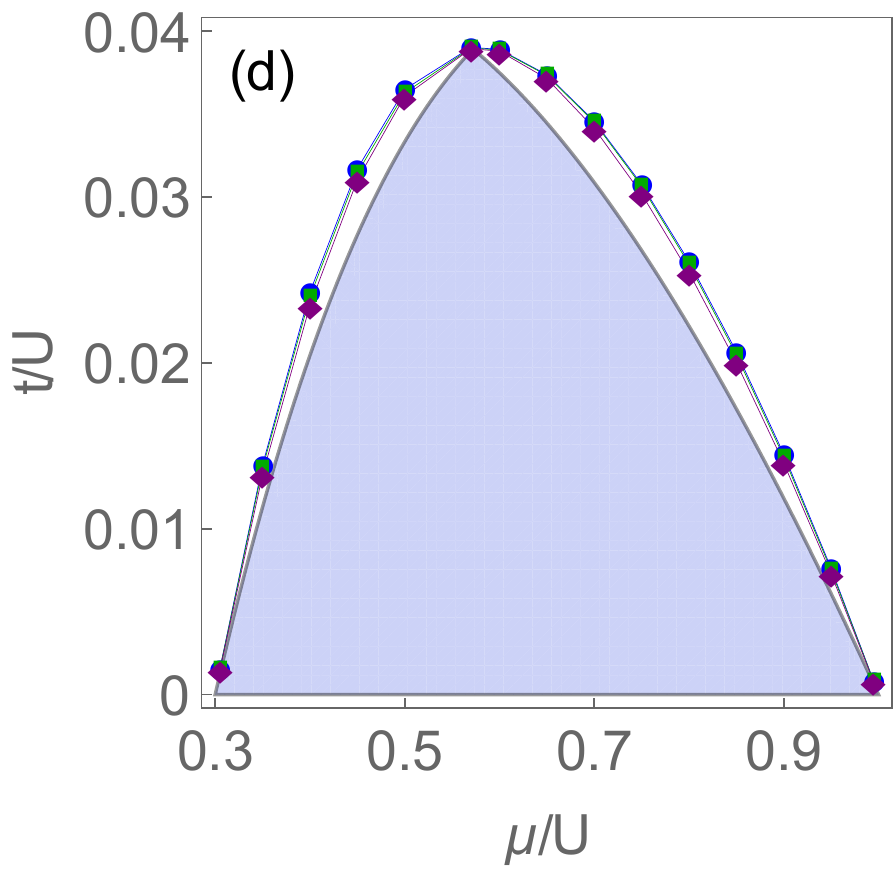}

\includegraphics[scale=0.5,trim=21 0 0 0]{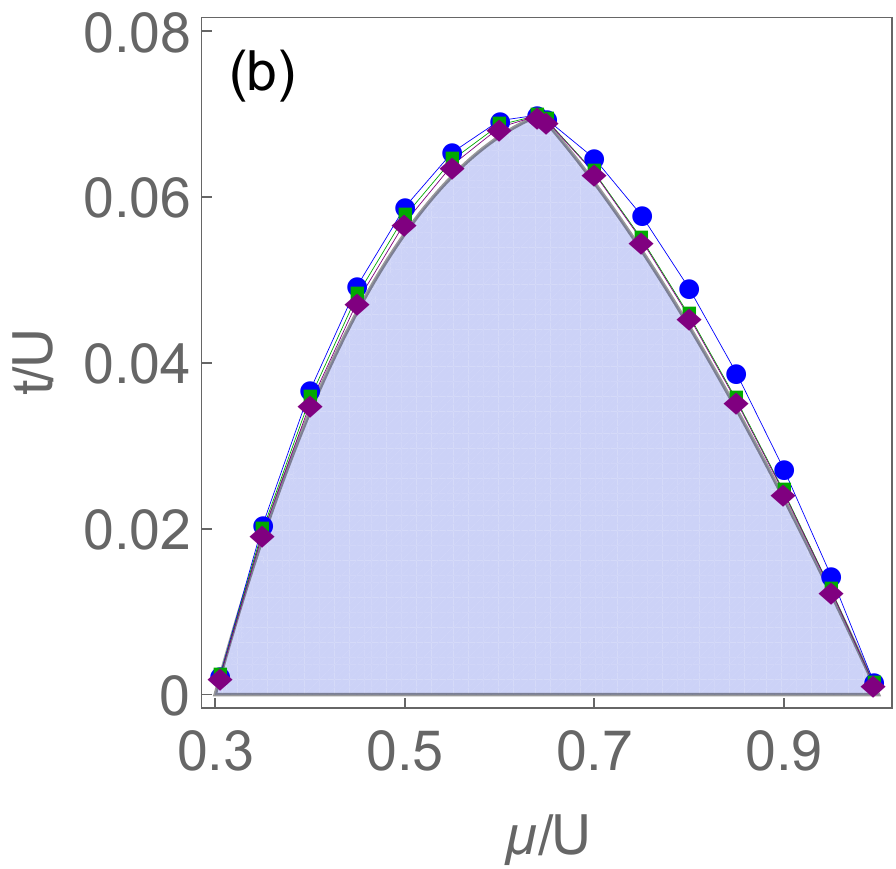}
\includegraphics[scale=0.5,trim=16 0 0 0,clip=true]{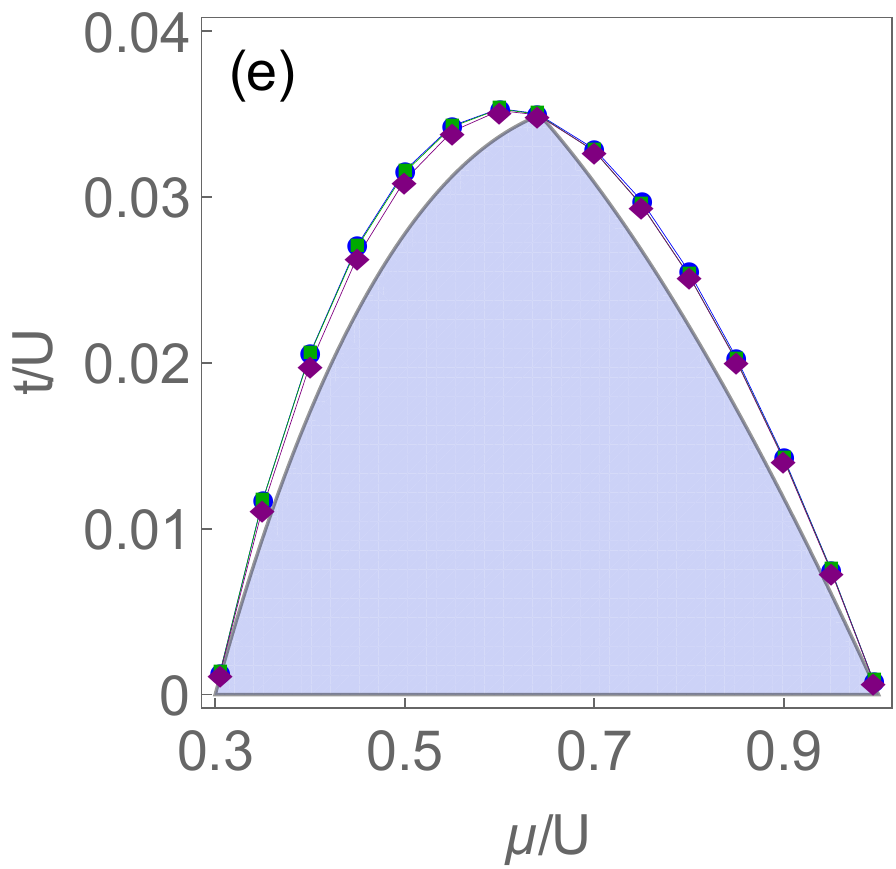}

\includegraphics[scale=0.5,trim=21 0 0 0]{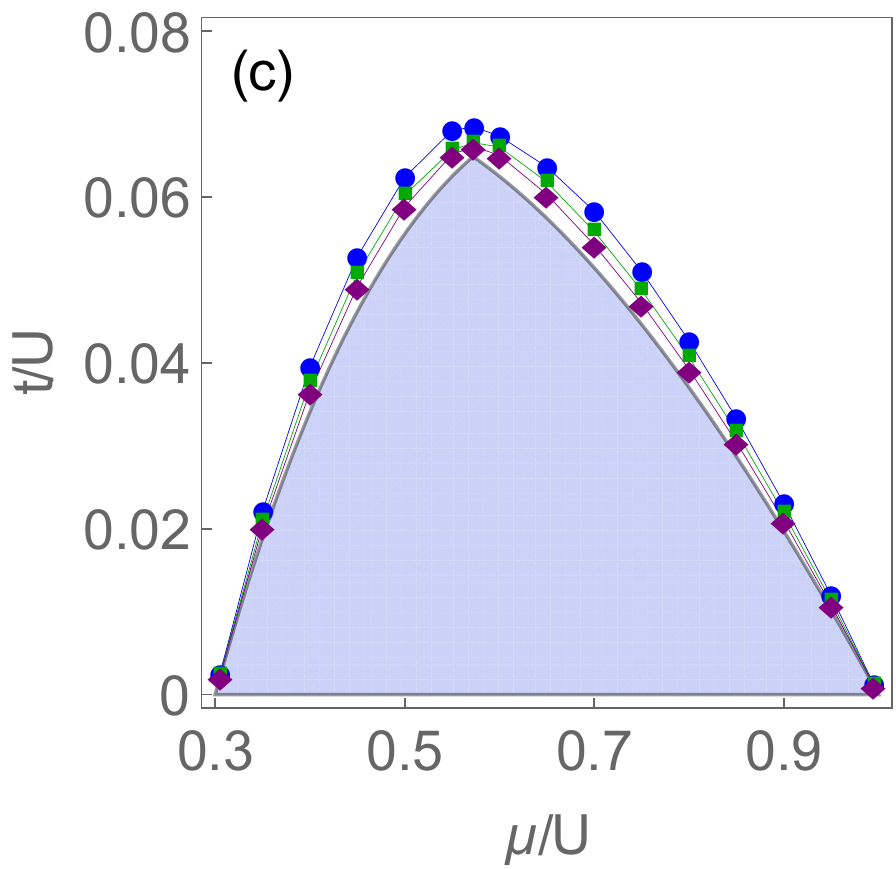}
\includegraphics[scale=0.5,trim=16 0 0 0,clip=true]{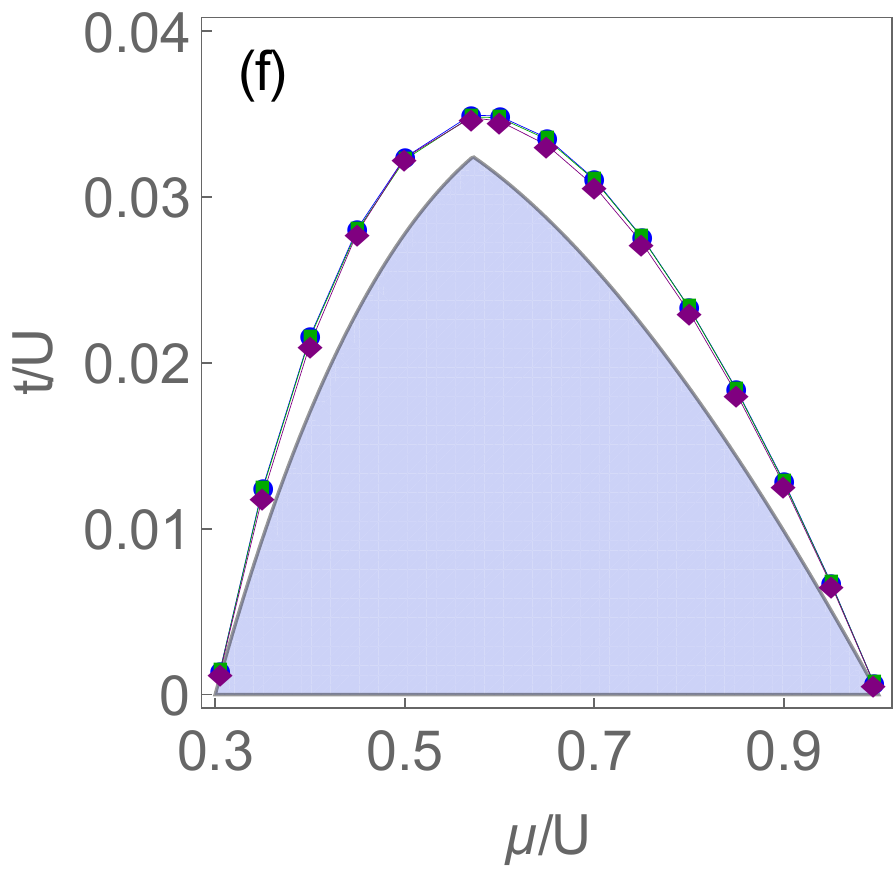}
\caption{Comparison of the numerical and theoretical boundaries of the Mott lobes for the Hamiltonians a), d) $H_1$, b), e) $H_2$ and c), f) $H_3$ in 1D (first column) and 2D (second column). The numerical results for 1D were obtained for $L=100,1000,10000$ while in 2D for $L=40,50,80$. Theoretical boundaries (with shaded area) are based on the analysis of the random variables characterizing spectrum of the matrix $\mathcal{R}$ (conditions in Eqs. (\ref{eq:Mott_1D}) and (\ref{eq:Mott_2D}) in the thermodynamic limit.}\label{fig:Mott}
\end{figure}
%
Consider first the MI phase and determination of its borders (MI lobes).  
We study the 1D systems of different lengths $L=100,1000,10000$ and the 2D system of size $L=40,50,80$ with open boundary conditions. This allows for semi-analytical expressions for the MI borders in the thermodynamic limit. In this limit  the properties of the system do not depend on the density of the impurities \cite{Mering2011pra} thus  we conveniently choose $p=0.5$. 
 We also choose $\gamma=0.3, \alpha=0.1$ in the model. Those parameters are, as mentioned above, determined by the boson-fermion interaction strength. For too strong interactions one would have to consider higher bands \cite{Mering2011pra,Jurgensen12} not included in our model. 

First, we find the boundary of the MI phase analyzing the spectrum of the matrix $\mathcal R$. In the 1D case $\mathcal R$ is a tridiagonal matrix with the following random off-diagonal upper ($X_i^{+}$) and lower ($X_i^{-}$) elements
\begin{eqnarray}\label{eq:Xpm_def}
 X_i^{\pm}&=&[1+\alpha(\Omega_i+\Omega_{i\pm 1})]\\
 &&\times\left(\frac{\bar{n}_i+1}{
\bar{n}_i-\mu+\bar \gamma \omega_i}-\frac{\bar{n}_i}{(\bar{n}_i-1)-\mu+\bar \gamma \omega_i}\right)\nonumber
\end{eqnarray}
and zeros elsewhere. As discussed by Mering and Fleishhauer \cite{Mering2008pra} (see also \cite{Buonsante2009pra}) in the thermodynamic limit the border of the fully incompressible Mott phase may be determined from {\it non-random} the situation, i.e., assuming that all sites $\omega_i$ are identical. It is due to the fact that it is the tunneling which kills MI and it is optimal at resonance, i.e., between identical sites. This argument holds also for our model with density induced tunneling terms. Then the random variables $X_i^{\pm}$ take the same value for each $i$ which allows us to estimate the spectrum of $\mathcal R$ using the formula for the tridiagonal Toeplitz matrix \cite{Bottcher2005}.
\begin{equation}\label{eq:Lambda_1D}
\Lambda_k=2 \sqrt{X_i^{+} X_{i}^{-}} \cos\left(\frac{k\pi}{L+1} \right),\quad k=1,\ldots,L.
\end{equation}
Now for large $L$ the condition (\ref{eq:Mottbound}) takes the form
\begin{equation}\label{eq:Mott_1D}
 2 \bar t \max\left[\sqrt{X_i^{+} X_{i}^{-}}\right]<1.
\end{equation}
Analogously for 2D we obtain
\begin{equation}\label{eq:Mott_2D}
 2 \bar t \max\left[\sqrt{X_{i,j}^{+} X_{i,j}^{-}}+\sqrt{Y_{i,j}^{+} Y_{i,j}^{-}}\right]<1,
\end{equation}
where $X_{i,j}^{\pm}, Y_{i,j}^{\pm}$ are given by expressions similar to (\ref{eq:Xpm_def}) (see Appendix \ref{app:RandomMatrices} for a more detailed derivation).
In Fig. \ref{fig:Mott} we compare the theoretical boundary of the Mott lobes for the Hamiltonians $H_1, H_2, H_3$ with the one obtained for finite systems of sizes $L=100,1000,10000$ (1D) and $L=40,80,100$ (2D). Note that for the binary disorder Mott lobes with non-integer average filling equal to $m+(1-p), m=0,\ldots$ (number of free sites) appear between the standard Mott lobes \cite{Buonsante2009pra} as for $m<\bar \mu<m+\bar \gamma$ the local number of bosons is site-dependent, i.e.,
\begin{eqnarray}
&&m< \bar \mu < \bar n_i < 1+\bar \mu < 1+m+\bar \gamma,\quad \text{for }\omega_i=0\\
&&m-\bar \gamma <\bar \mu-\bar\gamma < \bar n_i < 1+\bar\mu-\bar\gamma < 1+m,\quad \text{for }\omega_i=1,\nonumber
\end{eqnarray}
which gives $n_i=1+m$ in the first case and $n_i=m$ in the second.

\begin{figure}[!ht]
\includegraphics[scale=0.5]{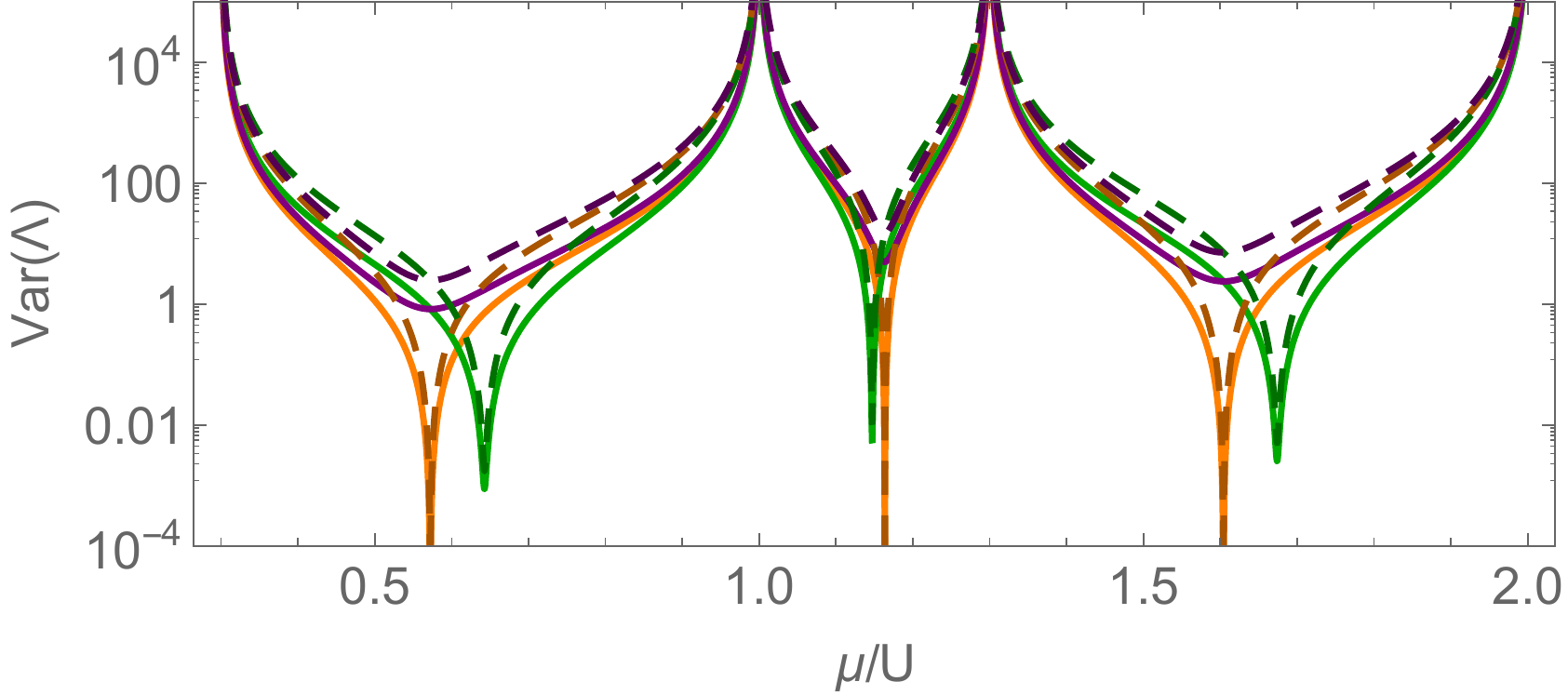}
\caption{Variance of the eigenvalues of $\mathcal R$ for 1D (dashed lines) and 2D systems (continuous lines) described by the Hamiltonians $H_1$ (orange), $H_2$ (green) and $H_3$ (purple).} \label{fig:spectra}
\end{figure}

In order to assess whether in any of the three analyzed cases a direct MI-SF transition is possible or if the system always passes through an intermediate BG phase we analyse the properties of the spectrum of $\mathcal R$.
As indicated in Eqs. (\ref{eq:Mott_1D}) and (\ref{eq:Mott_2D}) the most relevant eigenvalues stemming from homogeneous region are $\sqrt{X_i^+X_i^-}$ for any $i$ in that region. For large enough region the cosine term in (\ref{eq:Lambda_1D}) may be approximated as unity.   
It turns out that for the Hamiltonian $H_1$ for each choice of the parameter $\gamma$  there exists $\mu$ for which the variance of the random variables $\sqrt{X_i^+X_i^-}$, ${\rm Var}(\Lambda)$ vanishes. This value of $\mu$ corresponds to the tip of the MI lobe. In other words, at this point, the matrix $\mathcal R$  is exactly Toeplitz and  not random.  We compare the variances of the largest eigenvalues of $\mathcal R$, corresponding to large homogeneous regions, for the three Hamiltonians in 1D and 2D in Fig.~\ref{fig:spectra}.
Clearly, for $H_2$ and $H_3$ the variance of the eigenvalues never becomes zero. Moreover, the minimum variance of the eigenvalues for $H_3$ is much larger than for $H_2$. This leads us to the conclusion that $H_1$ may have triple points and a direct MI-SF phase transition whenever ${\rm Var}(\Lambda)$  vanishes. 
The addition of the disorder in the tunneling term in $H_2$  removes the triple points. In this case at least a thin layer of the BG phase surrounds the Mott lobe everywhere. In the last situation, in which the disorder in the tunneling is independent from the random chemical potential, the area of the BG should be much wider despite the same strength of the disorder.

\begin{figure}[ht!]
\includegraphics[scale=1]{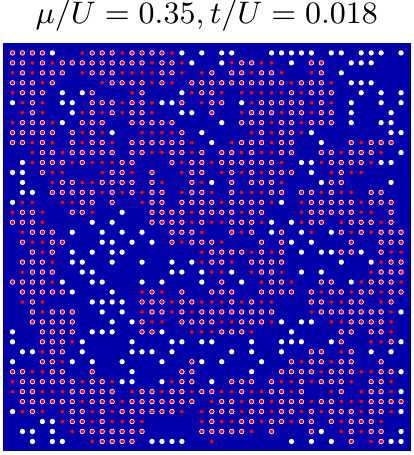}\includegraphics[scale=1]{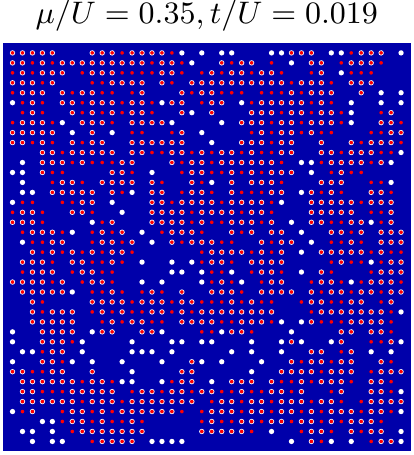}

\includegraphics[scale=1]{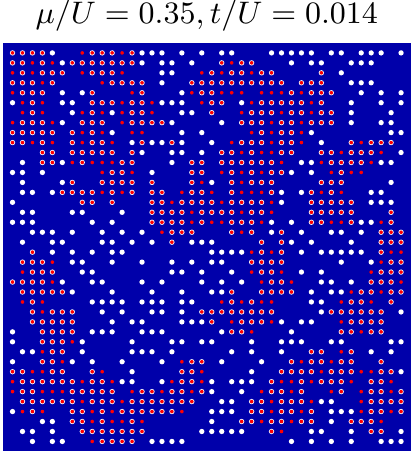}\includegraphics[scale=1]{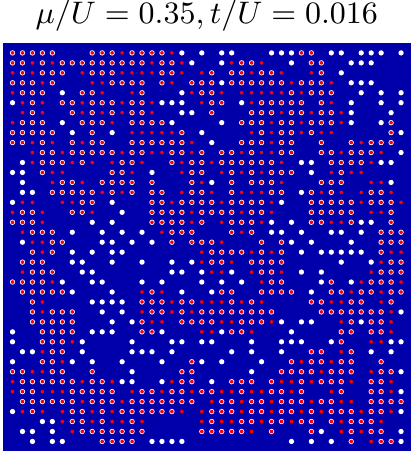}

\includegraphics[scale=1]{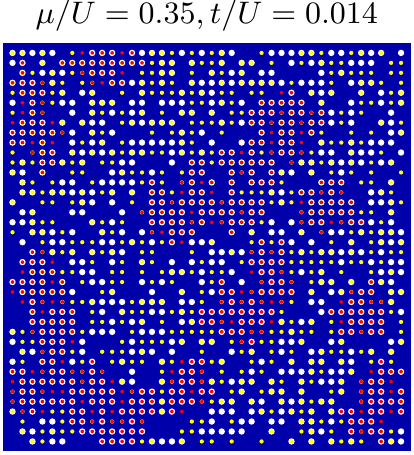}\includegraphics[scale=1]{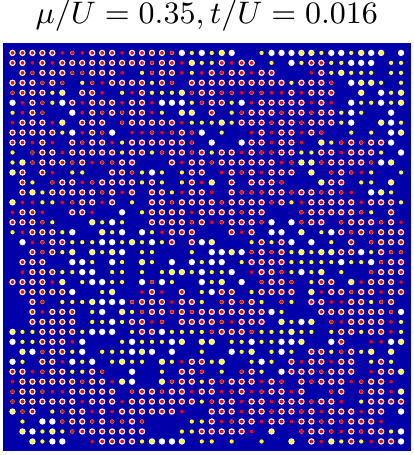}
\caption{Typical distribution of the SF clusters, i.e., the regions of the lattice with non-zero local occupation by the ``superfluid'' particles, in the systems described by Hamiltonians $H_1$, $H_2$, $H_3$. In the first column the clusters do not percolate, hence the phase is identified as the Bose-glass. In the second column, the clusters begin to percolate and the SF phase emerges.}\label{fig:num_distribution}
\end{figure}

We now determine the boundary of the BG by finding the distribution of $\av{n_i}$ in the lattice as in Eq. (\ref{eq:avnum}). We compute the energy and the number of particles in the SF clusters using the redefined modes and the information about their population. The total number of particles which is adjusted to match the chemical potential provides an estimate of the condensate fraction.
In Fig. \ref{fig:num_distribution} we show a typical distribution of the SF clusters in the regime of the BG and in the point of the transition to the SF phase in 2D.

%
%
%

\begin{figure}[ht!]
\includegraphics[scale=0.3]{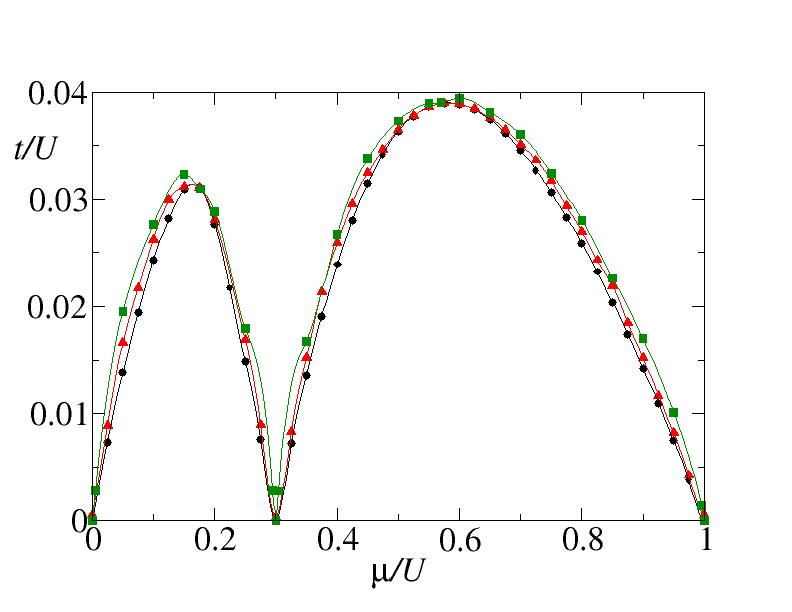}\\
\includegraphics[scale=0.3]{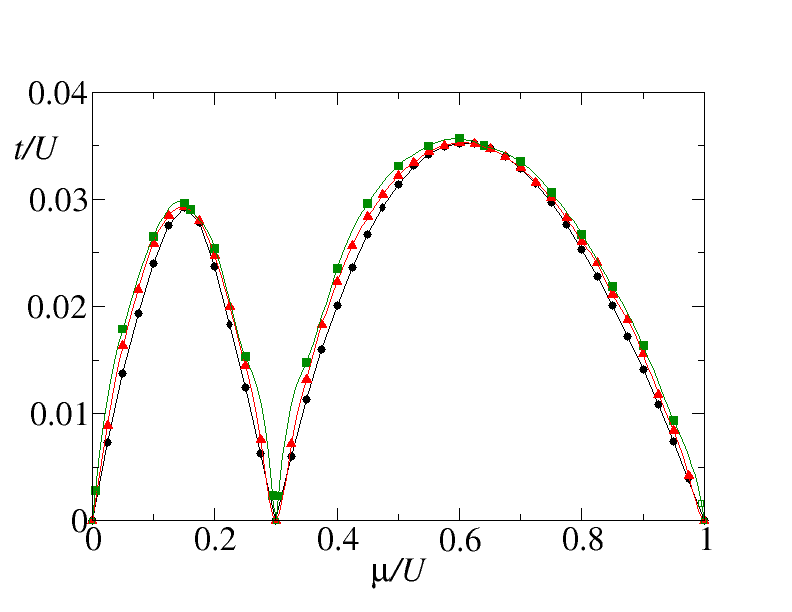}\\
\includegraphics[scale=0.3]{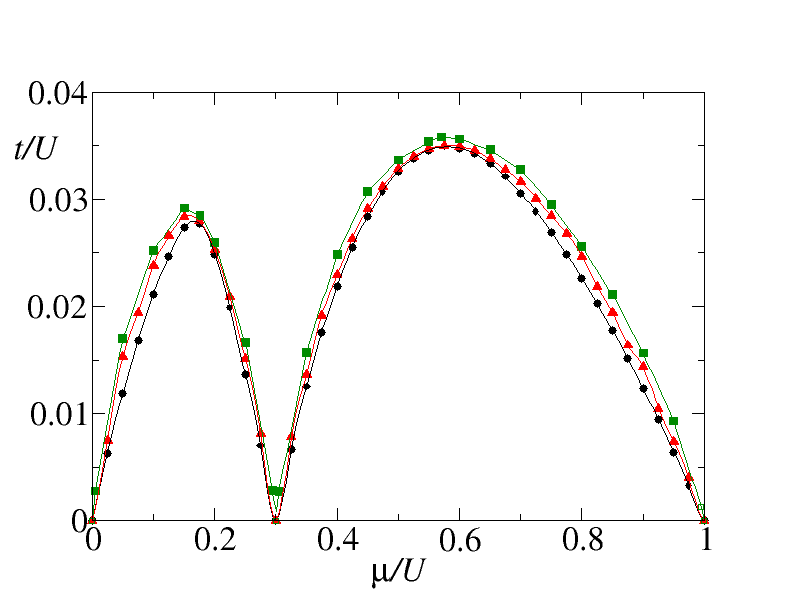}
\caption{Bose-glass regions for Hamiltonians $H_1$, $H_2$, $H_3$ (from top to bottom). Black line (with circles) give the MI-BG border, red line (triangles) corresponds to Gutzwiller ansatz BG-SF border obtained using superfluid fraction threshold, green line (squares) estimates the same border using HF percolation approach. Note that the latter is consistently higher than the SF prediction.}
\label{fig:glass2D}
\end{figure}

In Fig.~\ref{fig:glass2D} we compare the phase diagrams for the Hamiltonians $H_1, H_2, H_3$ obtained using the percolation approach with the BG-SF border coming from the Gutzwiller ansatz as explained in the previous sections. Observe that the percolation approach gives a consistently larger BG region, notably for $H_3$, although in all the cases we find that the BG regions are quite small in 2D for our choice of parameters. This is in contrast to the 1D situation with prominent BG regions \cite{Buonsante2009pra} for $H_1$ and also for $H_2$ and $H_3$ (not shown). We do not present the results for 1D since the mean field does not give reasonable quantitative predictions in this case.  

\section{Summary and discussion}\label{sec:summary}
We studied the Bose-Hubbard model with binary disorder obtained by pinning a secondary (fermionic) type of atoms in the optical potential. We revealed the following differences between the models describing such system: i) the model with disorder entering only in the potential term admits a direct MI-SF transition, which is not in contradiction with the theorem of inclusions; ii) in the model with disorder given by the same random variable affecting the tunneling and the chemical potential the transition to the SF phase goes always through an intermediate Bose-glass phase; iii) the disorder in the tunneling and in the potential given by independent random variables makes the intermediate Bose-glass region a bit thicker then the correlated disorder of point ii). Still the BG region in 2D for the given type and strength of disorder yields only a rather narrow slip around the Mott lobes.

Moreover we introduced a new type of mean-field approach for disordered systems, which combines the local mean-field approach with a simple ``Hartree-Fock-like'' procedure. Its advantage stems from the fact that maintaining the simplicity of the local mean-field approach it allows one to bring some spatial correlations into the description.

\acknowledgments 
The authors wish to thank R. Chhajlany and R. Augusiak for discussion.
The work of O.D.\ and J.Z.\ has been supported by Polish National Science Centre within project No. DEC-2012/04/A/ST2/00088. M.L. and J.S. acknowledge the financial support from Spanish Government Grant TOQATA (FIS2008-01236), EU IP SIQS, EU STREP EQuaM, and ERC Advanced Grants QUAGATUA and OSYRIS. J.S. acknowledges the support of fundaci\'o Catalunya - La Pedrera.
M.\L. acknowledges support of the Polish National Science Center by means of project no. 2013/08/T/ST2/00112 for the PhD thesis, and a research grant DEC- 2011/01/N/ST2/02549. M.\L. also acknowledges a special stipend of Smoluchowski Scientific Consortium "Matter Energy Future". Numerical simulations were performed thanks to the PL-Grid project: contract number: POIG.02.03.00-00-007/08-00 and at Deszno supercomputer (IF UJ) obtained in the framework of the Polish Innovation Economy Operational Program (POIG.02.01.00-12-023/08).

\appendix

\section{Estimation of the spectrum of the random matrix $\mathcal R$}\label{app:RandomMatrices}
We study the 1D case as the first step in the iterative procedure which allows us to estimate the spectrum of the matrix $\mathcal{R}$ for any dimension.

The structure of the matrix $\mathcal{R}$ for 1D is the following:
\begin{equation}\label{R_1D}
\mathcal{R}^{\rm 1D}=\left(\begin{array}{ccccc}
   0 & X_{1}^{+} &  &  & \\
   X_{2}^{-} & 0 & X_{2}^{+} &  & \\
   & X_{3}^{-} & 0 & X_{3}^{+} & \\
     &  & \ddots & \ddots & \ddots
   \end{array}\right),
\end{equation}
where
\begin{eqnarray}\label{a2}
 X_i^{\pm}&=&[1+\alpha(\Omega_i+\Omega_i\pm 1)]\nonumber\\
 &&\times\left(\frac{\bar{n}+1}{
\bar{n}-\bar \mu+\bar \gamma \omega_i}-\frac{\bar{n}}{(\bar{n}-1)-\bar\mu+\bar\gamma \omega_i}\right),
\end{eqnarray}
and $\mu-\gamma \omega_i < \bar n_i < 1+\mu-\gamma \omega_i$.
The elements in the off-diagonal bands form a set of identical random variables, therefore to estimate the spectrum we treat them as identical elements in the tridiagonal Toeplitz matrix of dimension $L\times L$ and evaluate the spectrum as
\begin{equation}\label{app:Lambda_1D}
\Lambda_k=2 \sqrt{X_i^{+} X_{i+1}^{-}} \cos\left( \frac{k\pi}{L+1} \right),\quad k=1,\ldots,L.
\end{equation}
Note that the random variable $\Lambda_k$ does not depend on the choice of $i$, as the random variables $X_i^{\pm}$ have identical distributions. The index is written explicitly only to specify how many independent $\omega$'s in (\ref{a2}) should be taken into account.  The maximum of $\Lambda_k$ is achieved for $k=L$ and is equal to the maximum value the random variable $2 \sqrt{X_i^{+} X_{i+1}^{-}}$ can take. Having the spectrum (\ref{app:Lambda_1D}) one can also estimate the variance of the maximal eigenvalues of $\mathcal R$, as
\begin{equation}
{\rm Var}[\lambda({\mathcal R^{\rm 1D}})]=4 \av{X_i^{+} X_{i+1}^{-}}-4 \av{\sqrt{X_i^{+} X_{i+1}^{-}}}^2.
\end{equation}
%

In the 2D case the random matrix $\mathcal R$ has the following block structure:
\begin{equation}\label{R_2D}
\mathcal{R}^{\rm 2D}=\left(\begin{array}{ccccc}
\mathcal{R}_1^{\rm 1D} & \mathcal{D}_1^{+} &  &  & \\
\mathcal{D}_2^{-} & \mathcal{R}_2^{\rm 1D} & \mathcal{D}_2^{+} &  & \\
    & \mathcal{D}_3^{-} & \mathcal{R}_2^{\rm 1D} & \mathcal{D}_3^{+} & \\
     &  & \ddots & \ddots & \ddots
   \end{array}\right),
\end{equation}
where $\mathcal{R}_i^{\rm 1D}$ is the matrix (\ref{R_1D}) for the $i$th row of the lattice,
\begin{equation}\label{R_i_1D}
\mathcal{R}_i^{\rm 1D}=\left(\begin{array}{ccccc}
   0 & X_{i,1}^{+} &  &  & \\
   X_{i,2}^{-} & 0 & X_{i,2}^{+} &  & \\
   & X_{i,3}^{-} & 0 & X_{i,3}^{+} & \\
     &  & \ddots & \ddots & \ddots
   \end{array}\right),
\end{equation}
with
\begin{eqnarray}
 X_{i,j}^{\pm}&=&[1+\alpha(\Omega_{i,j}+\Omega_{i,j\pm 1})]\\
&&\times \left(\frac{\bar{n}_{i,j}+1}{U
\bar{n}_{i,j}-\mu+\gamma \omega_{i,j}}-\frac{\bar{n}_{i,j}}{U (\bar{n}_{i,j}-1)-\mu+\gamma \omega_{i,j}}\right).\nonumber
\end{eqnarray}
The off-diagonal block-band contains diagonal matrices of the form
\begin{equation}\label{D}
\mathcal{D}_i^{\pm}=\left(\begin{array}{ccccc}
   Y_{i,1}^{\pm} &  &  &  & \\
    & Y_{i,2}^{\pm} &  &  & \\
   &  & Y_{i,3}^{\pm} &  & \\
     &  &  & \ddots & 
   \end{array}\right),
\end{equation}
with
\begin{eqnarray}
 Y_{i,j}^{\pm}&=&[1+\alpha(\Omega_{i,j}+\Omega_{i\pm 1,j})]\\
 &&\times\left(\frac{\bar{n}_{i,j}+1}{
\bar{n}_{i,j}-\bar \mu+\bar \gamma \omega_{i,j}}-\frac{\bar{n}_{i,j}}{ (\bar{n}_{i,j}-1)-\bar\mu+\bar\gamma \omega_{i,j}}\right).\nonumber
\end{eqnarray}
We again estimate the eigenvalues of $\mathcal{R}^{\rm 2D}$ using the formula for the spectrum of the $L\times L$ Toeplitz matrix analogously to the 1D case, only now the bands consist of of identical blocks of size $L\times L$. The random variables estimating the spectrum then read:
\begin{eqnarray}\label{eq:Lambda_2D}
\Lambda_{k,l}&=&\Lambda_k^{(i)}+2\sqrt{\lambda(D_i^{+})\lambda(D_{i+1}^{-})}\cos\left(\frac{l\pi}{N+1}\right)\nonumber\\
&=&2\sqrt{X_{i,j}^{+} X_{i,j+1}^{-}} \cos\left( \frac{k\pi}{L+1} \right)\nonumber\\
&&+2\sqrt{Y_{i,j}^{+}Y_{i+1,j}^{-}}\cos\left(\frac{l\pi}{L+1}\right),\nonumber\\
&&\hspace{0.5\columnwidth} k,l=1,\ldots,L
\end{eqnarray}
Note that the random variable $\Lambda_{k,l}$ again does not depend on the choice of $i$, as the random variables $X_i^{\pm}$ and $Y_i^{\pm}$ have identical distributions and once more we write it explicitly only to specify independent $\omega$s that should be taken for the calculation. The maximum of $\Lambda_{k,l}$ is achieved for $k=L, l=L$ and is equal to the maximum value of the sum $2 \sqrt{X_{i,j}^{+} X_{i,j+1}^{-}}+2 \sqrt{Y_{i,j}^{+} Y_{i+1,j}^{-}}$. The variance of the eigenvalues of $\mathcal R$ in the 2D case can be estimated similarly to the 1D case.

The procedure can be further iterated for higher dimensions.


\end{document}